\documentclass[%
 reprint,
 superscriptaddress,
 onecolumn,
 amsmath,
 amssymb,
 10pt,
 aip,
 jcp,
 citeautoscript,
 notitlepage,
 nofootinbib
]{revtex4-1}

\usepackage{graphicx}
\usepackage{dcolumn}
\usepackage{array}
\usepackage{bm}
\usepackage[%
  colorlinks=true,
  allcolors=blue,
  breaklinks=true
]{hyperref}
\usepackage{breakcites}
\usepackage{gensymb}
\usepackage[utf8]{inputenc}

\begin{document}

\title{Memetic Algorithms for Ligand Expulsion from Protein Cavities}

\author{J. Rydzewski}
\email[Email: ]{jr@fizyka.umk.pl}
\affiliation{Institute of Physics, Faculty of Physics, Astronomy and
Informatics, Nicolaus Copernicus University, Grudziadzka 5, 87--100 Torun,
Poland}

\author{W. Nowak}
\affiliation{Institute of Physics, Faculty of Physics, Astronomy and
Informatics, Nicolaus Copernicus University, Grudziadzka 5, 87--100 Torun,
Poland}


\begin{abstract}
\vspace*{0.05cm}
\noindent\fbox{%
  \parbox{0.82\textwidth}{%
    This is the updated version of work published previously in 
    \textit{J.~Chem.~Phys.} \textbf{143}, 124101
    (2016)\cite{jcp}.
  }%
}%
\vspace*{0.4cm}

Ligand diffusion through proteins is a fundamental process governing biological
signaling and enzymatic catalysis. The complex topology of protein tunnels
results in difficulties with computing ligand escape pathways by standard
molecular dynamics (MD) simulations. Here, two novel methods for searching
of ligand exit pathways and cavity exploration are proposed: memory random
acceleration MD (mRAMD), and memetic algorithms (MA). In mRAMD, finding exit
pathways is based on a non-Markovian biasing that is introduced to optimize
the unbinding force. In MA, hybrid learning protocols are exploited to predict
optimal ligand exit paths. The methods are tested on three proteins with
increasing complexity of tunnels: M2 muscarinic receptor, nitrile hydratase,
and cytochrome P450cam. In these cases, the proposed methods outperform
standard techniques that are used currently to find ligand egress pathways.
The proposed approach is general and appropriate for accelerated transport of
an object through a network of protein tunnels.
\end{abstract}

\maketitle


\section{Introduction}
Ligand recognition is one of the most critical steps in biological signaling
\cite{bhalla1999emergent}. To pass a signal, a ligand usually binds to a
specific receptor site which may be exposed on the receptor surface or buried
within the receptor matrix. Ligand residence time is of crucial
importance in regulatory processes\cite{tummino2008residence}. Entrance, binding
and egress processes involve complex migration through the protein
tunnels or ligand accessible cavities. Properties of these pathways determine
rate of signaling. Uncovering the distributions of transport routes is important
not only in understanding mechanisms of signal transduction, but also in
enzymatic catalysis, molecular diseases, and drug design\cite{bhat2000cellulases}.
Ligand dissociation may be studied computationally, but classical molecular
dynamics (MD) suffers from the low probability of crossing energy barriers
which kinetically entrap ligands in the protein matrix. To facilitate the
crossing, and in turn, to increase rare conformational event probability,
numerous enhanced MD methods have been developed so far, i.e.,
locally enhanced sampling (LES)\cite{elber1990enhanced,nowak1991reaction},
targeted MD\cite{schlitter1994targeted}, steered MD (SMD)\cite{kosztin1999unbinding},
and supervised MD\cite{sabbadin2014supervised}.
For a review of enhanced MD methods see, for example,
Ref.~\onlinecite{johnston2014beyond}.

The techniques proposed here, mRAMD and MA, may be considered as an extension of 
SMD, with the time-dependent direction of the unbinding force. It is worth to mention 
that SMD in its original form is limited in sampling optimal ligand egress paths, 
because the unbinding force given by:
\begin{equation}
  {\bf F}=-\frac{k}{2}\nabla_{\bf r}\left[vt-({\bf r}-{\bf r}_0)\cdot{\bf n}\right]^2,
\end{equation}
where $k$ is a force constant, $v$ is the constant unbinding velocity, ${\bf r}$ 
and ${\bf r}_0$ are the current and the initial positions of a pulled atom, and
${\bf n}$ is the unbinding direction; is kept constant during the simulation, so 
the sampled path is limited to a straight line. Also, the unbinding direction 
must be assumed a priori which is a severe drawback of SMD for protein tunnels
that are nonlinear.

A solution for this problem was found by L\"{u}demann et al.\cite{ludemann2000substrates} 
who introduced a method called random acceleration MD (RAMD), in which the unbinding direction is
modified when the traveling ligand meets any steric obstacle
\cite{wang2007chromophore,wang2009ligand}. This method made the enforced egress of 
camphor from cytochrome P450cam possible with no 
prior knowledge of exit tunnels\cite{ludemann2000substrates}. Despite its
popularity\cite{klvana2009pathways,li2011exploring,kalyaanamoorthy2012exploring}, 
RAMD has one drawback: a trade-off between artificial perturbation of the protein 
structure caused by the enforced ligand unbinding, and the low probability of 
sampling ligand unbinding events. In other words, many long trajectories have to be
computed to get a reasonable 
statistics\cite{ludemann2000substrates,vashisth2008ligand}.

\begin{figure}[h]
\includegraphics[width=0.7\textwidth]{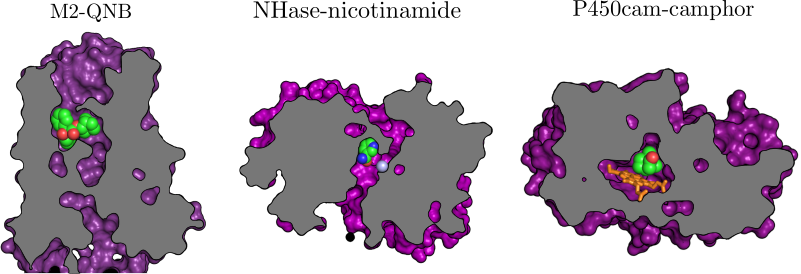}
\caption{Structures of the ligand-protein complexes studied in this article
  shown in increasing complexity of protein tunnels: M2-QNB, NHase-nicotinamide 
  with cobalt (light blue) in the catalytic center of NHase, P450cam-camphor with 
  heme shown (orange).}
\label{fig:1}
\end{figure}

Here, two new methods for studying ligand transport through complex protein 
channels and tunnels are introduced: mRAMD and MA. Both find plausible unbinding 
paths from buried protein binding sites in the following test systems: M2 muscarinic receptor 
(M2), nitrile hydratase (NHase), and cytochrome P450cam (P450cam). The last
system, P450cam, was studied in the original RAMD
work\cite{ludemann2000substrates} (Fig.~\ref{fig:1}).

\section{Methods}


\subsection{Non-Markovian Variant of Random Acceleration MD}

We start by introducing RAMD which is an extension of SMD that speeds up 
dissociation kinetics by few orders of magnitude, and allows to find 
various probable dissociation routes. Also no prior knowledge of an exit 
tunnel or channel is required. The RAMD protocol is the following:
\begin{enumerate}
  \item The direction ${\bf\hat r}$ of the unbinding force acting on the center of mass 
    of the ligand is assigned randomly, ${\bf f}=f{\bf\hat r}$, where $f$ is a 
    constant magnitude of the unbinding force.
	\item The unbinding force is maintained for a predetermined number of simulation 
    steps, $m$. The ligand is expected to move with a velocity exceeding the 
    threshold velocity given by $v_t=r_t/m\Delta t$, where $\Delta t$ is the time 
    step and $r_t$ is a specified minimum distance that the ligand passes before 
    the unbinding direction changes. If this condition is not fulfilled, the
    unbinding direction is reassigned randomly.
\end{enumerate}

Finding optimal values for all parameters in RAMD ($r_t$, $f$, $m$) for a 
ligand-protein complex is far from trivial. Vashisth and
Abrams\cite{vashisth2008ligand} considered it to be a drawback of RAMD as 
the adoped force constant $f$ should not be too high. This resulted in the 
percentage of successful dissociation events at about
19\%-41\%\cite{vashisth2008ligand}.

We introduced modifications to the standard RAMD method. First, 
we added an additional heuristic constraint to decrease a high 
number of unsuccessful ligand unbinding trajectories from a receptor:
\begin{enumerate}
\setcounter{enumi}{2}
	\item Next random unbinding direction is chosen if the distance traveled 
    by the ligand during the current $i$th time interval $d(m\Delta t_i)$ of 
    the simulation is smaller than the distance traveled in the previous 
    interval $d(m\Delta t_{i-1})$. This usually happens when some steric 
    obstacle emerges along the sampled unbinding route.
\end{enumerate}

Second, we introduced a variant of RAMD with a non-Markovian dependency added
(mRAMD). It was developed for finding the most probable egress pathways. 
The dependence on the previous simulated unbinding trajectories was added to 
mRAMD as an additional positive reinforcement. This process, inspired by swarm
optimization methods\cite{dorigo1996ant}, leads to the initial random probing of the 
protein tunnels, but gradually while more trajectories are sampled, the ligand 
experiences not only the stochastic force in a random direction $f_0\bm{\hat r}$, 
but also the force directed to dense regions of conformational space sampled 
by the previous trajectories $f_1 \bm{\hat k}$, i.e., $\bm{f}=f_0 \bm{\hat r}+f_1 \bm{\hat k}$.

The initial distribution of the ligand conformations in the protein tunnel 
is an important factor. A reasonably good guess can be obtained by running an exploratory 
LES\cite{elber1990enhanced} simulation, or by simply collecting the previous 
conformations gradually from mRAMD simulations. Clearly, $f_0$ should be 
larger than $f_1$, otherwise the resulting unbinding pathways would be 
constraint to the initial distribution of conformations, and their diversity 
would be limited.

During mRAMD simulations some paths may exhibit steric clashes and kinetic traps.
To eliminate these problem we applied the following scheme to the
positive reinforcement protocol in mRAMD:
\begin{enumerate}
	\item To eliminate rarely visited paths, the reinforcement continuously
    decreases during the simulation according to $\rho_i=\rho_i(1-q)$, 
    where $\rho_i$ is the density of the $i$th trajectory and $q$ is a 
    damping factor (set to 0.01).
	\item Density of previous conformations is averaged. The Shepard 
    approximation algorithm\cite{shepard1968two} is used with KD-trees for finding
    nearest neighbors. This reduces the complexity of the Shepard 
    method to $O(N\log N)$, where $N$ is the number of interpolated neighbors
    \cite{friedman1977algorithm}. We used the Liszka
    kernel\cite{liszka1984interpolation} (for details see Supporting
    Material).
\end{enumerate}


\subsection{Finding Unbinding Pathways with Non-Convex Optimization}

Finding and exploring ligand unbinding pathways from proteins may be
formulated in terms of an optimization problem. Solving such a problem is then
equivalent to finding an extreme of a postulated multivariate scoring
function. Here, out of many scoring-based metaheuristics suitable for such 
studies we used memetic algorithms (MAs) which are a good choice for non-convex
scoring functions\cite{goldberg2006genetic,michalewicz1996genetic}. MAs 
involve an iterative process of learning (i.e., adapting to the scoring function). 
Ligand conformations were represented by their center of mass positions within a predefined 
sampling radius, $r_s$, centered at the ligand. 

The optimization scheme in MAs is the following:
\begin{enumerate}

  \item Initially, the ligand conformations are chosen by sampling center of
  mass positions inside the sampling radius, $r_s$, centered at the current 
  ligand conformation whose dynamics is given by the MD simulation.

  \item The score $f^{(i)}$ of the $i$th conformation is calculated based on 
  effective interaction energy between the corresponding ligand and 
  the protein. In our implementation of MAs, we considered effective 
  interaction energy consisting of four terms:

  \begin{equation}
  \begin{split}
  \label{eq:1}
  f^{(i)} & =\alpha_{v}\sum_{k<l}\left({A_{kl}\over r_{kl}^{12}}-{B_{kl}\over r_{kl}^6}\right)\\
          & +\alpha_{h}\sum_{k<l}\left({C_{kl}\over r_{kl}^{12}}-{D_{kl}\over r_{kl}^{10}}\right)\\
          & +\alpha_{e}\sum_{k<l} {q_k q_l \over \epsilon(r_{kl}) r_{kl}}\\
          & +\alpha_{s}\sum_{k<l} (S_k V_l + S_k V_l)\exp{(-r_{kl}^2/2\sigma^2)},
  \end{split}
  \end{equation}

  where the $\alpha_{(\cdot)}$ coefficients on the right-hand side are
  empirically determined using linear regression from a set of 
  ligand-protein complexes\cite{morris1998automated,goodsell1990automated}. 
  The summations are over ligand indices $k$ and protein indices $l$. The 
  first term is the Lennard-Jones 12--6 potential. The second term is hydrogen
  bond energy modeled by the Lennard-Jones 12--10 potential. $A$, $B$, 
  $C$, and $D$ are matrices calculated to mimic the depth of the 
  Lennard-Jones potential well and equilibrium bond distances for homogeneous 
  pairs of atoms. The third term describes the Coulombic potential, where $q$ 
  is the charge of a given atom. The distance-dependent dielectric variable 
  is modeled by a sigmoid:

  \begin{equation}
    \epsilon(r)=a+{b\over {1+k\exp(-\lambda br)}}
  \end{equation}
    
  where $b=\epsilon_0-a$ and $\epsilon_0=78.4$ (dielectric constant in water in
  25 C), $a=-8.5525$, $k=7.7839$, and $\lambda=0.003627~\text{\AA}^{-1}$. The last term 
  in Eq.~\ref{eq:1} represents desolvation energy. In this work we used 
  partial atomic volumes $V$ and solvation coefficients $S$ from  Stouten et al
  \cite{stouten1993effective}. In other words, the last term describes to what 
  extent the protein buries the ligand in its interior\cite{wesson1992atomic}.

  \item Selection mechanism depends on scores of the ligand conformations. The 
  lower effective interaction energy of a ligand, the higher the probability 
  of surviving to a next epoch. We used the roulette selection scheme:

  \begin{equation}
    \label{sel}
    p_i={f^{(i)}\over{\sum_{j=1}^n f^{(j)}}},
  \end{equation}

  where $f^{(i)}$ is the score of the $i$th conformation, and $p_i$ is the resulting 
  probability of being selected to the next epoch.

  \item Randomly selected ligand conformations undergo perturbations: 
  combination and a Cauchy deviation. Conformations created during 
  combination replace their precursors in the sampling set. The Cauchy 
  deviation is performed by deviating the center of mass position by a number 
  sampled from a Cauchy distribution.

  \item Additional optimization procedure is applied in MAs, namely, a local 
  search that leads to a faster convergence during the learning phase. We 
  used two stochastic local searches: stochastic hill climbing (SHC)
  \cite{mitchell1993will,skalak1994prototype}, and the Solis-Wets method (SW)
  \cite{solis1981minimization}. In SHC the current ligand conformation is
  replaced only if a stochastically perturbed neighboring ligand has lower
  effective interaction energy. For this the algorithm proposed by Forest 
  and Mitchell was used\cite{mitchell1993will}. In the case of ligand-protein 
  complexes, a random neighbor is created by the Cauchy deviation of the current 
  ligand. In SW, the sampling domain is dynamically adjusted to increase the 
  success rate of finding a better solution
  \cite{hart1994adaptive,morris1998automated,goodsell1990automated}.

  \item The steps described in 2--5 comprise an epoch. After multiple epochs the 
  convergence is reached, and the ligand conformation with lowest effective 
  interaction energy is taken as a next milestone. In other words, the unbinding 
  proceeds in the direction of the selected conformation during the MD
  simulation. After a predefined number of steps the optimization procedure is 
  repeated starting from the next ligand conformation.

\end{enumerate}

We tested three variants of MAs: MA with SHC (MA-SHC), MA with SW (MA-SW),
and MA without a local search employed (MA).


\section{Simulation Protocol}


\subsection{Optimization Parameters} 

In the test systems studied with MAs, the size of the learning set was 20. 
Ligand conformations were deviated by perturbing their center of mass
positions. This was applied to a randomly chosen conformation with the 
probability of 0.02. The mean and spread of Cauchy distribution were set to 
0 and 1, respectively. The optimization process was stopped after 20 epochs. 
The combination rate was set to 0.8. The local searches were applied to 
a randomly selected conformation with the probability of 0.67. The sampling radius,
$r_s$ from which conformations were sampled was 8~\AA~in M2-QNB, 15~\AA~in
NHase-NCA, and 10~\AA~P450cam-camphor.

For the parameters used in the mRAMD simulations see Tabs. I-III, and for
details regarding the MA simulations of the ligand-protein complexes, see 
Supporting Material.


\subsection{Implementation}

The main advantage of our program is a capability to compute ligand unbinding
pathways during MD simulations. We used NAMD2.9 code\cite{phillips2005scalable}
to perform MD simulations with the CHARMM27 force field\cite{brooks1983charmm}. 
The communication between NAMD and the implemented program is done via NAMD's
feature called external program forces which is an interface to calculate
biasing forces. The methods presented in this study were implemented in the 
C++11 programming language\cite{stroustrup1995c++} using boost
\cite{karlsson2005beyond}. The code is available on Github
(\url{https://github.com/jakryd/maze-namd}).


\section{Results and Discussion}

\begin{figure}[h]
\centering
\includegraphics[width=0.5\textwidth]{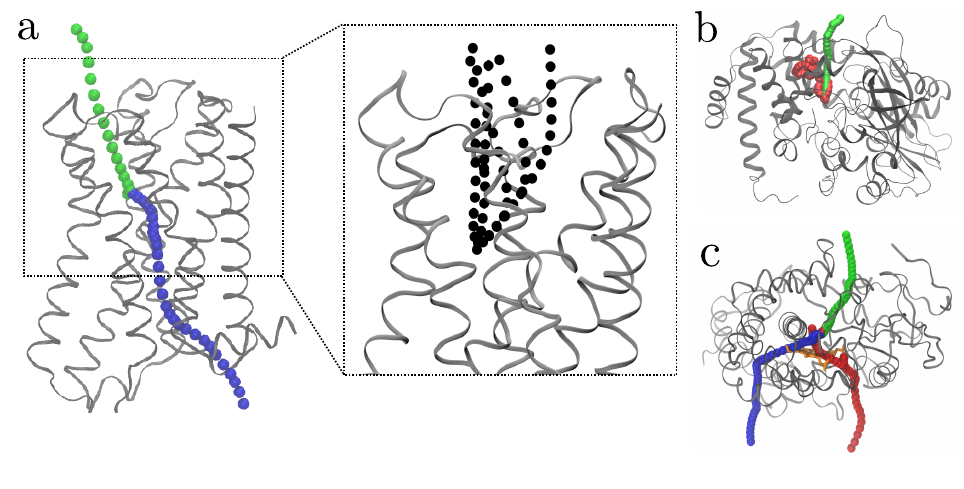}
\caption{Example ligand unbinding reaction pathways of the studied receptors. 
  (a) M2-QNB complex: PW1 (green), PW2 (blue); example trajectories of PW1 
  (black) are given in the inset, (b) NHase-nicotinamide complex: PW (green), a
  trajectory that did not dissociate during a RAMD simulation (red), (c) P450cam-camphor 
  complex: PW1 (blue), PW2 (green), PWC (red).}
\label{fig:2}
\end{figure}

We studied ligand unbinding paths in the following model systems: M2
muscarinic receptor, nitrile hydratase, and cytochrome P450cam. In these
complexes the channels accessible to ligands show increasing complexity
(Fig.~\ref{fig:1}). We used mRAMD and MAs to sample ligand unbinding pathways and compared the
results with RAMD. The resulting unbinding paths of are schematically 
presented in Fig.~\ref{fig:2}. In order to assess the efficiency of methods, a simple 
statistics was collected for each model system, for instance, the success rate of
dissociation, defined as a ratio of the number of successful ligand exits from the
protein tunnel to the number of all computational trials for a given
method\cite{vashisth2008ligand,carlsson2006unbinding}. We note that unbinding 
times in simulations are short because unbinding forces used to enforce 
dissociation events for each complex were high.

\begin{table*}[t]
\centering
\caption{Characteristics of the unbinding pathways in the M2-QNB complex. 
  The results of the simulations for the force constant -- $f$~[kcal/mol~\AA], $m$ -- 
  the frequency of choosing the next unbinding direction (in steps), $N$ - the number of 
  simulations, $N_{p}$ - the number of successful unbinding trajectories for a 
  given pathway. Values for unbinding time $t$~[ps], work $w$~[kcal/mol], and distance 
  $d$~[\AA] are shown.}
\begin{tabular}{lcccp{0.5cm}cccccccccc}
  \hline
  \noalign{\smallskip}
  Method & Pathway & $f$ & $m$ & $N$ & $N_{p}$ Min. $t$ & Avg. $t$ & Max. $t$ & Min. $d$ & Avg. $d$ & Max. $d$ & Min. $w$ & Avg. $w$ & Max. $w$ \\
  \hline
  RAMD      & PW1 & 5  & 10 & 20 & 9  & 0.86 & 1.11 & 1.35 & 26.56 & 30.15 & 37.30 & 75.6  & 100.4 & 126.3 \\
  RAMD      & PW2 & 5  & 10 & 20 & 3  & 0.77 & 1.57 & 2.34 & 15.92 & 43.86 & 64.14 & 78.3  & 118.7 & 179.9 \\
  RAMD      & PW1 & 10 & 10 & 20 & 5  & 0.45 & 0.72 & 0.99 & 24.84 & 38.39 & 65.73 & 193.0 & 260.4 & 408.5 \\
  RAMD      & PW2 & 10 & 10 & 20 & 6  & 0.98 & 1.24 & 1.44 & 43.84 & 60.57 & 79.80 & 326.2 & 442.6 & 595.9 \\
  mRAMD     & PW1 & 5  & 10 & 9  & 6  & 0.75 & 1.11 & 1.65 & 21.15 & 29.16 & 36.59 & 71.7  & 93.9  & 115.6 \\
  mRAMD-LES & PW1 & 5  & 10 & 10 & 10 & 0.64 & 0.9  & 1.14 & 18.18 & 25.49 & 32.77 & 63.7  & 89.7  & 118.7 \\
  MA        & PW1 & 5  & 50 & 20 & 12 & 0.55 & 0.71 & 0.97 & 18.26 & 24.47 & 31.94 & 70.8  & 111.4 & 148.5 \\
  MA-SHC    & PW1 & 5  & 50 & 10 & 9  & 0.51 & 0.79 & 1.18 & 17.39 & 26.58 & 39.08 & 83.4  & 117.6 & 182.9 \\
  MA-SW     & PW1 & 5  & 30 & 10 & 8  & 0.57 & 0.85 & 1.25 & 17.76 & 28.57 & 41.02 & 80.0  & 126.1 & 164.4 \\
  \hline
\end{tabular}
\end{table*}


\subsection{QNB in the M2 Muscarinic Receptor}

In M2, the antagonist is buried 20~\AA~inside a regular cylindrical cavity 
(Fig.~\ref{fig:1}). We found two QNB dissociation pathways. The shapes of the 
PW1 egress trajectories (Fig.~\ref{fig:2}) are similar in all the methods
studied, which perhaps means that it is a natural way for QNB to reach the exterior 
of M2. However, the average distance of diffusion, as indicated by the parameter 
Avg. $d$ in Tab. I, is the shortest for the path predicted by mRAMD and MA (25~\AA).
For RAMD the ligand had to travel approximately 30~\AA~before reaching the
receptor exterior region. The M2 muscarinic receptor with the QNB ligand bound 
is in an inactive conformation which means that a hydrophobic gate consisting 
of 3 amino acids (LEU65, LEU114, ILE392) is tightly closed\cite{haga2012structure}. 
There are also theoretical studies showing that such a hydrophobic gate 
is connected with an activation of G-protein-coupled receptors\cite{yuan2014activation}. 
In M2, the hydrophobic gate is lying approximately 3~\AA~below the binding pocket 
of M2 and it blocks the PW2 egress pathways. The closed hydrophobic gate lowers
the probability of sampling PW2 in our calculations. To confirm this hypothesis,
we run multiple SMD simulations in the directions of exits identified by PW1 and PW2. 
The pulling forces needed to rupture the hydrophobic gate through PW2 were higher
by approximately 300~pN, comparing to the highest force resulted on PW1.
Moreover, in a recent paper by Kruse et al.\cite{kruse2012structure} authors 
studied QNB diffusion paths in the M3 muscarinic receptor which has a similar 
GPCR structure. Numerous very long MD simulations (nearly 25 microseconds) 
were used to estimate pathways for the spontaneous QNB association with the M3 receptor. 
The paths found in that paper are qualitatively similar to PW1 which suggests
that PW1 is preferable for QNB.

\begin{figure}
\includegraphics[width=\textwidth]{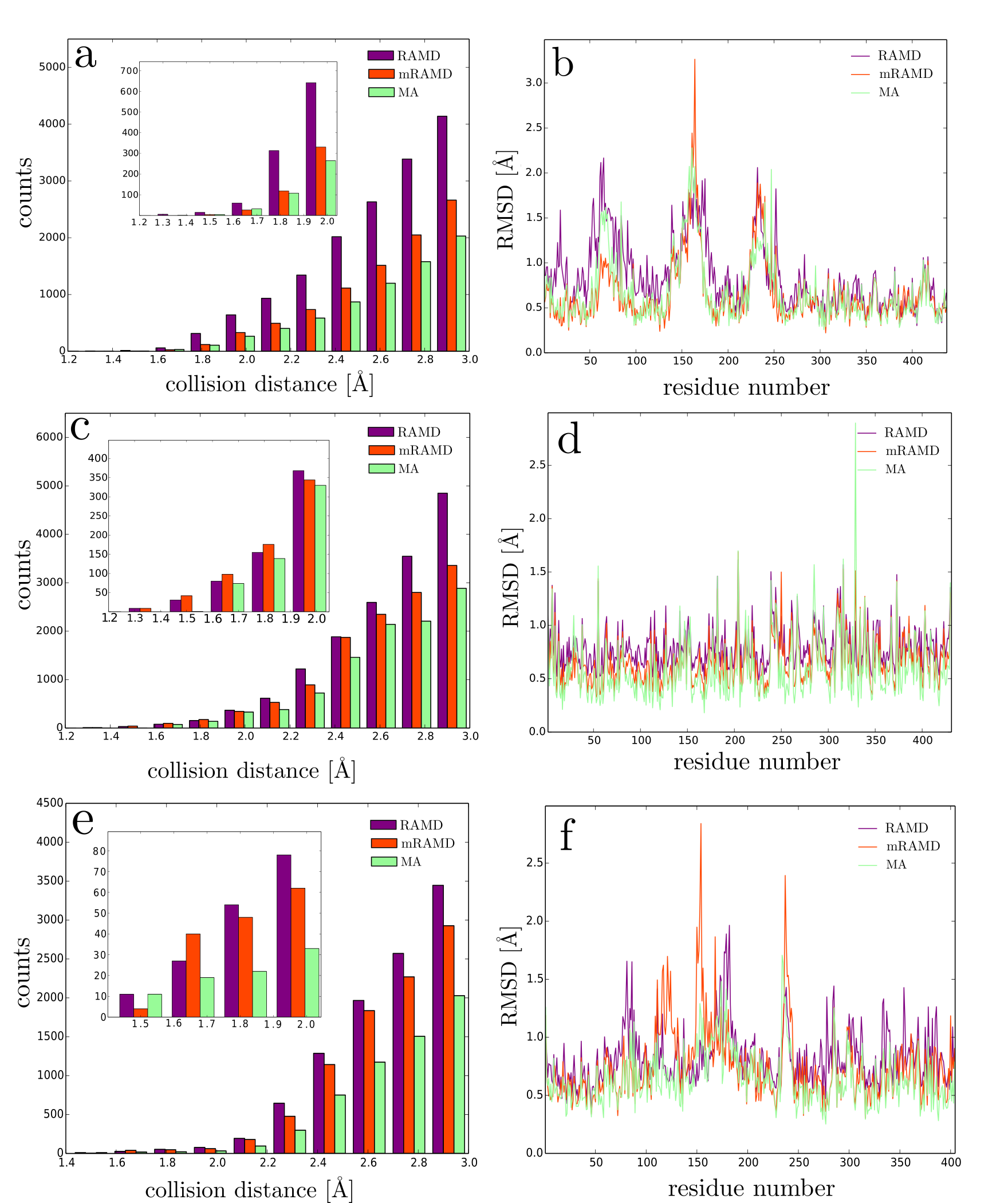}
\caption{Average RMSD per residue (b, d, f) and the collision statistics (a, c,
  e) of the M2-QNB (a, b), NHase-nicotinamide (c, d) and P450cam-camphor (e, f) 
  complexes from 10 randomly chosen trajectories. Insets (a, c, e) depict zoomed 
  histograms for the collision distance $<2$~\AA.}
\label{fig:3}
\end{figure}

We compared work done by unbinding force in the tested methods.
It is the lowest in MAs, and has a value of about 90 kcal/mol, while the
lowest value for Avg. $w$ in RAMD is 100 kcal/mol. Thus, we expect that the
perturbations of the receptor structure induced by the process of
ligand dissociation are smaller in mRAMD and MAs than it is in RAMD, which is another
advantage of the proposed methods (Fig.~\ref{fig:3}). It is worth noting that
Avg. $w$ for PW2 trajectories calculated by RAMD was as high as 442
kcal/mol (Tab. I).


\subsection{Non-Convex Optimization Performs better for Curved Nitrile Hydratase
Tunnel}

We calculated ligand exit paths for the nicotinamide-NHase system 
starting from the ligand buried approximately 40~\AA~beneath the protein surface. 
In contrast to the M2 receptor the channel is
curved, thus diffusion in this case is more challenging from the methodological
point of view. The methods found the same exit pathway, PW (Fig.~\ref{fig:2}b). It is in a
good agreement with our previous LES calculations for other
amides\cite{peplowski2008mechanical}. The results of the success rate and
pathway characteristics are presented in Tab. II.
mRAMD shows a moderate success rate for this system, 44\%. However, when
information on possible ligand conformations is obtained from LES,
the success rate rises up to 90\% (Tab. II). Clearly this suggests that LES
performs better the initial sampling of the protein interior due to lowered
potential energy barriers. 

MAs show an excellent success rate, despite the lack of the initial sampling:
MA - 90\% , MA-SHC - 100\%, and MA-SW - 100\%. The success rate of RAMD is low: 0\%-14\%. 
If the unbinding path is found, the mean unbinding time Avg. $t$ in RAMD is 5 times
higher than in mRAMD. It shows that the algorithms with a more advanced 
sampling scheme find optimal egress pathways. The lowest value for the work
performed during ligand unbinding is about 69 kcal/mol for MA-SW, and the largest 
value of 482 kcal/mol resulted from RAMD.

\begin{table*}[h]
\centering
\caption{Characteristics of the unbinding pathways in the NHase-NCA complex. 
  The results of the simulations for the force constant -- $f$~[kcal/mol~\AA], $m$ -- 
  the frequency of choosing the next unbinding direction (in steps), $N$ - the number of 
  simulations, $N_{p}$ - the number of successful unbinding trajectories for a 
  given pathway. Values for unbinding time $t$~[ps], work $w$~[kcal/mol], and distance 
  $d$~[\AA] are shown.}
\begin{tabular}{lcccp{0.8cm}cccccccccc}
  \hline
  \noalign{\smallskip}
  Method & Pathway & $f$ & $m$ & $N$ & $N_{p}$ Min. $t$ & Avg. $t$ & Max. $t$ & Min. $d$ & Avg. $d$ & Max. $d$ & Min. $w$ & Avg. $w$ & Max. $w$ \\
  \noalign{\smallskip}
  \hline
  RAMD      & PW & 10 & 50  &  14 & 2  & 5.35 & 5.75 & 6.15 & 49.72 & 117.48 & 185.24 & 247.9 & 482.0 & 716.1 \\
  RAMD      & PW & 15 & 30  &   7 & 0  & ---  & ---  & ---  & ---   & ---    &  ---   & ---   & ---   & ---   \\
  mRAMD     & PW & 10 & 100 &   9 & 4  & 1.15 & 3.71 & 6.00 & 38.96 & 77.64  & 153.78 & 121.9 & 242.1 & 473.3 \\
  mRAMD-LES & PW & 10 & 100 &  10 & 9  & 1.15 & 1.61 & 1.95 & 33.85 & 42.66  & 55.03  & 82.1  & 106.5 & 134.7 \\
  MA        & PW & 10 & 50  &  10 & 9  & 1.30 & 1.68 & 1.90 & 31.45 & 46.83  & 54.41  & 49.5  & 76.5  & 95.6  \\
  MA-SHC    & PW & 10 & 50  &  15 & 15 & 1.10 & 1.66 & 1.95 & 21.07 & 44.49  & 54.56  & 36.7  & 72.5  & 89.0  \\
  MA-SW     & PW & 10 & 50  &  10 & 10 & 1.55 & 1.63 & 1.80 & 32.57 & 43.33  & 49.18  & 44.0  & 69.5  & 77.4  \\
  \hline
\end{tabular}
\end{table*}


\subsection{Sampling Transient Tunnels in Cytochrome P450cam}

As the last and the most complex test case, cytochrome P450cam from
\textit{Pseudomonas Putida} was used (Fig.~\ref{fig:1}). During the P450cam 
activity camphor enters a distal pocket buried inside the enzyme and located 
close to the heme moiety. As previously said, L\"{u}demann et al. used RAMD to study possible 
camphor dissociation paths\cite{ludemann2000substrates}. After running
hundreds of simulations the authors found three distinct routes (1, 2, 3),
but for the path no. 2 three variants were observed. The results of our
RAMD calculations for P450cam agree with those obtained by
L\"{u}demann et al. we found three groups of pathways as well.
PW1 path (Fig.~\ref{fig:2}c) is identical to that presented in L\"{u}demann et al. 
The same refers to our PW2 and the pathway no. 2a. However, we
were not able to observe path no. 3. Instead we found the third pathway in P450cam,
which corresponds to so-called water channel (PWC, Fig.~\ref{fig:2}c). PWC
corresponds to a hydrophilic channel located near the heme propionic groups, and
was already suggested as the exit channel for the product, 5-hydroxy-camphor, which
is more hydrophilic than camphor\cite{poulos1986crystal,oprea1997identification}. 
There are lacking evidences that show the existence of PWC in P450s cytochromes 
in standard RAMD calculations\cite{schleinkofer2005mammalian,winn2002comparison,wade2004survey}. 
The function of PWC is not clear so far. It may have a role in the passage of water
and molecular oxygen. PWC opens towards the proximal side of the heme group,
which suggests, that it may have a role in the electron transport system
\cite{wade2004survey}. Alternatively, it may be involved in the transport of
protons through a network of ordered water molecules\cite{vohra2013dynamics}.

The detailed results on the pathways in cytochrome P450cam are presented in 
Tab. III. One can see that only 60\% of the RAMD trials led to camphor exit
($f=10$ kcal/mol\AA). If the force constant of $f=5$ kcal/mol\AA~was used 
in RAMD, the success rate was even lower (20\%). However, in the case of 
MAs proposed here, the success rates were much better. mRAMD gave the 80\% 
success rate, and mRAMD with LES provided 100\% (PW1 30\%, PW2 70\%). All the
variants of MA-based methods gave a 100\%
success rate. The majority of the methods predicts that PW2 dominates in 
the camphor transport in cytochrome P450cam, in accordance with the earlier 
report\cite{ludemann2000substrates}.

mRAMD needs 2.63 ps to find an exit for the same conditions as for RAMD (PW2, 
$f=5$ kcal/mol\AA), thus the total simulation time for mRAMD is
approximately 3 times shorter than in RAMD. A comparison of the
distance traveled by camphor in both cases (Avg. $d$, Tab. III) favors
mRAMD as well (43.8~\AA~and 25.53~\AA for RAMD and mRAMD, respectively). 
Interestingly, the shortest distance path PW2 was
predicted by the MA-SW method (15.44~\AA). This very short path was quickly
found (Avg. $t$, 1.53 ps) due to a variable sampling domain adopted in the
IA-SW algorithm. This feature is particularly advantageous when the diffusion
channel is very narrow, like in P450cam. All the other methods
needed numerous trials before a successful unbinding direction was
determined. Here, in MA-SW a larger sampling domains quickly brings
information that larger area is available for a ligand. The work Avg. $w$ (see
Tab. III) performed by unbinding force along PW2 is the smallest
one (66 kcal/mol). A relatively low work of 90 kcal/mol is needed to force
camphor through PW2 as predicted by mRAMD, but as much as
305 kcal/mol is required to run the ligand along PW1 by RAMD.

\begin{table*}[h]
\centering
\caption{Characteristics of the unbinding pathways in the P450cam-CAM complex. 
  The results of the simulations for the force constant -- $f$~[kcal/mol~\AA], $m$ -- 
  the frequency of choosing the next unbinding direction (in steps), $N$ - the number of 
  simulations, $N_{p}$ - the number of successful unbinding trajectories for a 
  given pathway. Values for unbinding time $t$~[ps], work $w$~[kcal/mol], and distance 
  $d$~[\AA] are shown.}
\begin{tabular}{lcccp{0.8cm}cccccccccc}
  \hline\noalign{\smallskip}
  Algorithm & Pathway & $f$ & $m$ & $N$ & $N_{p}$ & Min. $t$ & Avg. $t$ & Max. $t$ & Min. $d$ & Avg. $d$ & Max. $d$ & Min. $w$ & Avg. $w$ & Max. $w$ \\
  \noalign{\smallskip}
  \hline
  RAMD      & PW1 & 10 & 100 & 10 & 2 & 1.30 & 2.03 & 2.75 & 33.06 & 49.83 & 66.60 & 287.6 & 417.6 & 270.4 \\
  RAMD      & PW2 & 10 & 100 & 10 & 3 & 1.05 & 1.15 & 1.30 & 27.73 & 29.16 & 30.67 & 214.4 & 242.8 & 270.4 \\
  RAMD      & PWC & 10 & 100 & 10 & 1 & 2.05 & 2.05 & 2.05 & 39.96 & 39.96 & 39.96 & 305.2 & 305.2 & 305.2 \\
  RAMD      & PW2 & 5  & 100 & 10 & 2 & 2.10 & 7.30 & 12.5 & 25.83 & 43.88 & 61.92 & 114.1 & 178.3 & 242.5 \\
  mRAMD     & PW1 & 5  & 100 & 10 & 3 & 2.10 & 2.63 & 2.90 & 25.59 & 31.26 & 39.12 & 85.4  & 106.7 & 141.8 \\
  mRAMD     & PW2 & 5  & 100 & 10 & 5 & 1.00 & 1.87 & 2.85 & 18.16 & 25.53 & 34.55 & 68.1  & 92.8  & 130.1 \\
  mRAMD-LES & PW1 & 5  & 100 & 10 & 3 & 2.50 & 3.88 & 5.35 & 29.20 & 38.15 & 49.86 & 90.2  & 115.6 & 150.6 \\
  mRAMD-LES & PW2 & 5  & 100 & 10 & 7 & 3.30 & 3.90 & 4.30 & 28.96 & 36.82 & 48.09 & 72.9  & 119.5 & 156.7 \\
  MA        & PW1 & 5  & 100 & 10 & 2 & 1.70 & 2.65 & 3.60 & 24.65 & 32.50 & 40.34 & 109.9 & 130.3 & 150.8 \\
  MA        & PW2 & 5  & 100 & 10 & 8 & 1.00 & 1.61 & 2.60 & 20.59 & 24.25 & 36.89 & 92.8  & 109.8 & 152.6 \\
  MA-SHC    & PW1 & 5  & 100 & 10 & 1 & 3.10 & 3.10 & 3.10 & 47.95 & 47.95 & 47.95 & 187.9 & 187.9 & 187.9 \\
  MA-SHC    & PW2 & 5  & 100 & 10 & 6 & 1.45 & 2.90 & 4.30 & 22.00 & 43.84 & 55.51 & 97.5  & 174.9 & 215.6 \\
  MA-SHC    & PWC & 5  & 100 & 10 & 3 & 3.45 & 4.68 & 5.30 & 50.73 & 66.43 & 77.64 & 190.9 & 254.7 & 296.8 \\
  MA-SW     & PW1 & 5  & 100 & 10 & 1 & 2.10 & 2.10 & 2.10 & 25.72 & 25.72 & 25.72 & 117.1 & 117.1 & 117.1 \\
  MA-SW     & PW2 & 5  & 100 & 10 & 3 & 1.20 & 1.53 & 1.70 &  8.95 & 15.44 & 20.98 & 32.0  & 66.2  & 90.8  \\
  MA-SW     & PWC & 5  & 100 & 10 & 6 & 2.20 & 2.89 & 3.40 & 21.53 & 24.18 & 27.64 & 92.1  & 99.3  & 116.1 \\
  \hline
\end{tabular}
\end{table*}

\subsection{Summary}

\begin{table}[!htbp]
\centering
\caption{Summary of the success rates for the ligand-protein complexes studied.}
\begin{tabular}{lllll}
\hline
\noalign{\smallskip}
  \multicolumn{1}{c}{method} & \multicolumn{4}{c}{success rate}\\
  {} & M2 & NHase & P450cam & Avg.\\
\noalign{\smallskip}
\hline
\noalign{\smallskip}
RAMD      & 60  & 14  & 60  & 44.6 \\
mRAMD     & 66  & 44  & 80  & 63.3 \\
mRAMD-LES & 100 & 90  & 100 & 96.7 \\
MA        & 60  & 90  & 100 & 83.3 \\
MA-SHC    & 90  & 100 & 100 & 96.7 \\
MA-SW     & 80  & 100 & 100 & 93.3 \\
\hline
\end{tabular}
\end{table}

The summary of the success rates for all algorithms is presented in Tab. IV.
We calculated the ratio of a number of successful paths to the number of all exit 
simulation attempts. The results show that
the best average success rate (Avg. $SR$) of 96.7\% is offered by the MERA-LES
algorithm, but one needs to remember that the mRAMD-LES calculations were based on
the pre-calculated ligand conformations by LES. Since MAs do not require
similar pre-calculations, they are in some sense the most successful in our study. The
MA-SHC and MA-SW methods display successful rate of 96.7\% and 93.3\%,
respectively. The mRAMD algorithm has $SR$ of 63.3\% which is
much better than RAMD (44.3\%). Moreover, the new
algorithms presented here show that collision statistics and the RMSD values
generated during 10 randomly chosen trajectories for each system tested, are
lower in MA, comparing to RAMD (FIG. 4, TABLE V). It is worth noting that
despite the high percentage of successful dissociations, MA and mRAMD do not
introduce excessive artifacts during the sampling of protein conformations.

\section{Conclusions}

Molecular modeling in biology, biophysics or drug design often requires
computation of unbinding pathways within macromolecular matrices. The 
classical MD simulations can contribute to prediction of such processes, 
however, simulations of diffusion may be extremely time-consuming.

In this paper we have presented two types of new methods that alleviate
these problems and improve the search for ligand unbinding pathways 
substantially: mRAMD and MAs. These methods have been tested on three 
protein-ligand systems with increasing complexity of the channels: M2 muscarinic
receptor and QNB, nitrile hydratase and nicotinamide, and cytochrome P450 and 
camphor.

In mRAMD, a memory was added to RAMD. Calculations showed that this 
method may improve RAMD. Moreover, in our variant we used locally enhanced 
sampling\cite{elber1990enhanced} to calculate initial distribution of ligand
conformations within the protein matrices. The resulting method showed a high 
success rate (96\%). One should note that LES pre-calculations require 
additional, but reasonable computing time.

The other group of new methods is based on optimization. In MAs a scoring 
function based on ligand-protein effective interaction energy is used to
assess the unbinding direction. We showed that local searches may be a good
solution for further optimization. The success rate of MAs is 93-96\% and they 
do not require any pre-calculations, as opposed to mRAMD.


\begin{acknowledgments}
This research was supported in part (WN) by the NCN grant N N202 262038. 
We thank for computer time allocated in the Interdisciplinary Center for Modern
Technologies, NCU. We would like to thank Rafal Jakubowski for useful discussions, 
and Aleksander Balter and Magdalena Rydzewska for reading over the manuscript.
\end{acknowledgments}


\bibliography{mv1}


\end{document}